\documentclass[bibauthoryear]{aa}

\usepackage{amssymb}

\DeclareMathOperator\arctanh{arctanh}

\usepackage{graphicx}
\usepackage[varg]{txfonts}
\usepackage{amsmath}
\usepackage{natbib}
\usepackage{xcolor}
\usepackage[colorlinks=true,citecolor=blue,linkcolor=blue,urlcolor=blue,anchorcolor=blue]{hyperref}
\begin{document}

	\title{Discriminating power of milli-lensing observations for dark matter models}
	
	\author{Nick Loudas
		\inst{1,2}
		\and
		Vasiliki Pavlidou\inst{1,2}
		\and
		Carolina Casadio\inst{2}
		\and
		Konstantinos Tassis\inst{1,2}
	}
	
	\institute{
		University of Crete, Department of Physics \& Institute of
		Theoretical \& Computational Physics, 70013 Herakleio, Greece \\\email{nloudas@physics.uoc.gr; pavlidou@physics.uoc.gr}
		\and 
		Institute of Astrophysics,
		Foundation for Research and Technology-Hellas, 71110 Heraklion, Crete, Greece\\
	}
	
	\date{Received / accepted }
	
	% \abstract{}{}{}{}{} 
	% 5 {} token are mandatory
	\abstract
	% context heading (optional)
	% {} leave it empty if necessary  
	{The nature of dark matter (DM) is still under intense debate. Sub-galactic scales are particularly critical, as different, currently viable DM models make diverse predictions on the expected abundance and density profile of DM haloes on these scales.}
	% aims heading (mandatory)
	{We investigate the ability of sub-galactic DM haloes to act as strong lenses on background compact sources, producing gravitational lensing events on milli-arcsecond scales (milli-lenses), for different DM models. For each DM scenario, we explore whether a sample of $\sim 5000$ distant sources is sufficient to detect at least one milli-lens.}
	% methods heading (mandatory)
	{We develop a semi-analytical model to estimate the milli-lensing optical depth as a function of the source's redshift for various DM models. We employ the Press-Schechter formalism, as well as results from recent N-body simulations to compute the halo mass function, taking into account the appropriate spherically averaged density profile of haloes for each DM model. We treat the lensing system as a point-mass lens and invoke the effective surface mass density threshold to calculate the fraction of a halo that acts as a gravitational lens. We study three classes of dark matter models: cold DM, warm DM, and self-interacting DM.}
	% results heading (mandatory)
	{We find that haloes consisting of warm DM turn out to be optically thin for strong gravitational milli-lensing (zero expected lensing events). CDM haloes may produce lensing events depending on the steepness of the concentration-mass relation. Self-interacting DM haloes can efficiently act as gravitational milli-lenses only if haloes experience gravothermal collapse, resulting in highly dense central cores.     }
	% conclusions heading (optional), leave it empty if necessary 
	{}
	
	\keywords{galaxies: haloes -- cosmology: dark matter --
		gravitational lensing: strong --
		methods: semi-analytical
	}
	\authorrunning{N. Loudas et al.}
	\titlerunning{Discriminating power of milli-lensing observations for dark matter models} 
	
	\maketitle
	
	%-------------------------------------------------------------------
	
	\section{Introduction} \label{sec1}
	
	One of the most
	groundbreaking findings during the last century was the discovery of a mass excess in nearby galaxies that could not be explained by the  
	amount of ordinary (luminous) matter that was found to exist in those galactic systems. The observation of flat rotation curves in spiral galaxies
	\citep[e.g.,][]{Rubin1980,Bosma1980,Corbelli2000}, the discrepancy between observed velocity dispersion measurements and those predicted by the virial theorem, in elliptical
	galaxies and globular clusters \citep[e.g.,][]{Zwicky,Faber1976}, and the presence of collapsed structures at high redshift \citep[e.g.,][]{Gunn1972} were smoking guns for the existence of a new extraordinary form of matter in the Universe, called dark
	matter (DM), which neither emits nor absorbs radiation.
	The DM hypothesis has led to several correct predictions and has explained many observational discrepancies that had emerged in the past from the comparison of observational data with the first cosmological scenarios that were dust-radiation-only models. Despite the great success of the DM model, its nature remains unknown, making it one of the most fundamental unsolved questions in physics. 
	
	The most widely accepted scenario for the origin of DM
	is the so-called cold dark matter (CDM), a part of the standard $\Lambda$CDM cosmological model that has been remarkably
	successful in explaining the properties of a wide range of large-scale observations, including the accelerating expansion of the Universe \citep{Perlmutter1999}, the
	power spectrum of the Cosmic Microwave Background (CMB) \citep{Page2003}, and the observed abundances of different types
	of light nuclei \citep{Cyburt2016}. However, the $\Lambda$CDM paradigm still presents some discrepancies with observations, mostly at small scales. Such small-scale challenges include, among others,
	the 'cusp/core' problem, the missing satellites problem, the too-big-to-fail
	problem, and the angular momentum catastrophe
	(for a review see \citealt{Bullock2017}; see also \citealt{Perivolaropoulos2021}). 
	
	An appealing solution to those problems is to modify the intrinsic properties of DM particles. During the past few years, numerous
	DM alternatives and $\Lambda$CDM extensions have been proposed by several authors, with the purpose to address some of
	the $\Lambda$CDM challenges. One of the most promising DM alternatives is the warm dark matter (WDM) model \citep[e.g.,][]{Viel2005,Lovell2012}, where particles have rest mass on the order of a few keV, such as sterile neutrinos or thermal relics, that had non-negligible velocities at early
	times.
	Another very popular DM scenario is the self-interacting dark matter (SIDM) model where particles interact with each other \citep[e.g.,][]{Spergel2000} having non-negligible cross-sections, of the order of $\sim 1\, \mathrm{cm^2/g}$ \citep[e.g.,][]{Zavala2013}. Other more exotic DM alternatives include; ultra-light axion dark matter \citep{Schwabe2016}, dark atoms \citep[for a review see][]{Cline2021}, and fuzzy dark matter \citep[e.g.,][]{Kulkarni2020}.
	Although these latter models are not examined here, the toolkit we have developed can be straight-forwardly adapted to any DM model for which the redshift-dependent mass function and the density profile of haloes and sub-haloes can be calculated.
	
	The properties of the DM particle affect the formation of DM structures on all scales, their stability, as well as their evolution in time.  
	In addition, the fundamental attributes of DM particles modify the primordial power spectrum describing the initial overdensity seeds of cosmological structures. So, differences in the intrinsic DM particle properties between different models are expected to lead to measurable deviations in the resulting mass function of collapsed objects. For instance, models which include light particles, such as WDM, feature a sharp cutoff in the differential halo mass function below a critical mass scale, which depends on DM particle mass: the mildly relativistic velocities of WDM particles in the early Universe led to small-scale density fluctuations being washed out (free streaming) \citep[e.g.,][]{Melott1985,Viel2005}. The density profile of DM haloes also turns out to be noticeably different from model to model. For example, virialized haloes made of WDM particles typically have lower central densities with respect to CDM haloes of the same mass, by virtue of their generally later formation epochs (\citealp[e.g.,][]{Lovell2012}).  
	
	The study of DM haloes below sub-galactic scales turns out to be particularly crucial for the exploration of the nature of DM. Nevertheless, it is extremely challenging to detect such haloes directly, in order to measure their number density in the Universe and/or examine their internal structure, since they might not even form galaxies due to their small size. So, the only possible way to explore and study them is through
	gravitational effects. 
	
	One of the most promising methods of detecting sub-galactic DM haloes is strong gravitational lensing, where light
	that passes near a massive object (the lens) is being deflected, traveling a longer path than it would in the absence of the gravitational potential
	of the lens \citep[e.g.,][]{Weinberg:100595}. 
	As a result, when a compact background source (for instance, a radio loud quasar) emits radiation with the lens being in between the source and the observer and close enough to the line-of-sight, then the path of the light is affected
	strongly, resulting in the emergence of multiple images of the background source on sky
	with different magnifications \citep[e.g.,][]{Vegetti2012}, provided its projected surface density exceeds a threshold. This effect is commonly known as
	strong lensing \citep[see for example,][] {Wright2000}. 
	In the special case where the source displays intrinsic
	variability, observable time delays between the different images (pulses) may occur \citep{Zackrisson2010}.

	Gravitational lensing can be used to detect compact objects (COs) that could not be detected otherwise, such as primordial black holes (PBHs) or dense DM haloes.
	\cite{PressGunn1973} introduced the idea of assessing the cosmological abundance of COs through their strong gravitational lensing effect on distant background sources. They demonstrated that the cosmological mass density of COs can be constrained by deriving the fraction of lensed radio sources.
	Later on, \cite{Wilkinson2001} carried out a search for milli-lenses (gravitational-lensing images with milli-arcsec separations) in Very Long Baseline Interferometry (VLBI) observations of a sample of 300 compact radio sources,
	but no lensed systems in the mass range $\sim 10^6\,\mathrm{M_\odot}$ to $\sim10^8\,\mathrm{M_\odot}$ were found. Their negative result allowed them to place an upper limit $\Omega_{\mathrm{CO}} \lesssim 0.01$ (95\% confidence) on the cosmological density of COs in this mass range, concluding that the contribution of a primordial supermassive BHs population to the dark matter content of the Universe is negligible. The currently on-going Search for MIlli-LEnses (SMILE) project (\citealp{Casadio2021}) expands the search for milli-lenses in the range $\sim 10^6\,\mathrm{M_\odot}$ to $\sim 10^9\,\mathrm{M_\odot}$, to a complete sample of $\sim 5000$ radio-loud sources using VLBI data.   
	
	Motivated by the potential of the SMILE project, in this work we develop a novel method to exploit its upcoming results with the purpose to derive constraints on the nature of DM and discriminate between currently viable DM scenarios. Our approach is based on the concept of the lensing optical depth, representing the probability for an observed source to be gravitationally lensed by a foreground mass distribution. The prescription for the implementation of this method can be found in \cite{Zackrisson2007}. Recently, several authors have followed similar approaches to place limits on the abundance of primordial black holes, using Fast Radio Bursts (FRBs) \citep[e.g.,][]{Leung2022,Zhou2022frb,Zhou2022a}, Gamma-ray Bursts (GRBs) \citep[e.g.,][]{Kalantari2021}, afterglows of GRBs \citep[][]{Gao2022} and compact radio sources \citep[][]{Zhou2022}. Here, we pursue the possibility of sub-galactic DM haloes acting as gravitational mill-lenses.
	
	We derive the expected number of milli-lenses in the source sample of the SMILE project for various DM models by calculating the milli-lensing optical depth as a function of the source's redshift. This in turn depends on the halo mass function, as well as on the projected surface mass density. Both of these physical quantities have noticeable differences between various scenarios, and hence the milli-lensing optical depth exhibits differences between DM models.    
	
	The layout of this paper is as follows. In Sect. \ref{sec2} we describe our calculation of the milli-lensing optical depth. In Sect. \ref{sec3} we discuss the analytic descriptions we use for the structure of DM haloes for various cosmological DM scenarios, and their corresponding mass functions. In Sect. \ref{sec4} we present the results of our calculations, which we discuss in Sect. \ref{sec5}
	%--------------------------------------------------------------------
	
	\section{Lensing probabilities} \label{sec2}
	
	The principal result of any survey for lensing systems in the observable Universe is the number of confirmed lensed images in a complete sample of sources \citep[e.g.,][]{Myers2003,Browne2003}. To maximize the constraining power of this product, we have to connect it to theoretical models that predict the expectation value of lensing events taking into account the differences in the abundance and density profile of DM haloes between various DM models. 
	The most straightforward way to achieve this is to compute the lensing optical depth for any given DM scenario. 
	
	The lensing optical depth depends strongly on the mass function of gravitational lenses and on the surface density profile of each halo which essentially is related to the density profile. It also depends on the cosmology. 
	In this paper, we fix the cosmological parameters to be $H_0 = 100~h ~\mathrm{km ~s^{-1} ~Mpc^{-1}}$, $h=0.7$, $\Omega_m=0.3$, $\Omega_\Lambda=0.7$, $n=0.97$, $\delta_c(0) = 1.674$, and $\sigma_8=0.8$. The overall results, however, are not sensitive to small variations in these parameters.

	\subsection{Milli-lensing optical depth} \label{sec2.1}
	In order for our results to be applicable to the SMILE project, we are interested in lenses that produce multiple images with angular separation on the order of milli-arcseconds (milli-lenses). Thus, we focus on lenses of masses $(10^6-10^9) \,\mathrm{M_\odot}$. 
	For the calculation of the mill-lensing optical depth, we adopt the prescription of \cite{Zackrisson2007}. We treat the lens as a massive object of mass $M_l$ with an angular Einstein radius
	\begin{equation} \label{2.1}
		\beta_E = \sqrt{4\dfrac{GM_l}{c^2}\dfrac{D_{ls}}{D_{ol}D_{os}}},
	\end{equation} 
	where $D_{os},\,D_{ls}, \text{and}\, D_{ol}$ are the angular-diameter distances from the observer to the source, from the lens
	to the source, and from the observer to lens, respectively, with the lens being at redshift $z$ while the source is located at redshift $z_s$. $D_{AB}$ can be written as
	\begin{equation} \label{2.3}
		D_{AB}(z_A,z_B) = \dfrac{c}{1+z_B} \int_{z_A}^{z_B}\dfrac{dz}{H(z)},
	\end{equation}
	where $H(z)$ is the Hubble parameter, \begin{equation}
		\left(\dfrac{H(z)}{H_0}\right)^2 
		%= (1+z)^{-2} \left(\dfrac{dz}{d(H_0t)}\right)^2
		= \Omega_{m}(1+z)^3 + \Omega_\Lambda,
		\label{3.1}
	\end{equation}
	with $\Omega_m,~\Omega_\Lambda$ referring to the present values of the density parameters for matter and dark energy, respectively.
	
	The milli-lensing optical depth for a source at redshift $z_s$ is given by
	\begin{equation} \label{2.4}
		\tau(z_s) = \int_{0}^{z_s}\left|\dfrac{cdt}{dz}\right| dz\int_{10^6 \mathrm{M_\odot}}^{10^9 \mathrm{M_\odot}}\sigma(M_l,z,z_s)\dfrac{dn(M_l,z,z_s)}{dM_l} dM_l,
	\end{equation}
	where $\sigma(M_l,z,z_s)$ is the lensing (effective) cross-section 
	\begin{equation} \label{2.5}
		\sigma(M_l,z,z_s) \equiv \pi \beta_{E}^2 D^2_{ol} = \dfrac{4\pi G M_l}{c^2} \dfrac{D_{ol}D_{ls}}{D_{os}},
	\end{equation}
	and $dn/dM_l$ is the differential lens mass function 
	\begin{equation} \label{2.6}
		\dfrac{dn(M_l,z,z_s)}{dM_l} = \dfrac{dM(M_l,z,z_s)}{dM_l}\dfrac{dn(M,z)}{dM}.
	\end{equation}
	In Eq. \eqref{2.6}, $dn/dM$ is the differential halo mass function (see \S \ref{sec3.1}). We caution the reader of the two different masses entering Eq. \eqref{2.6}: the lens mass $M_l$, and the halo mass $M$. These two are not in general the same because only the part of the halo in which the projected surface mass density exceeds the critical strong lensing threshold can act as a gravitational lens. The critical surface density value for a source at redshift $z_s$ undergoing strong gravitational lensing by a foreground DM halo (lens) at redshift $z$ is
	\begin{equation} \label{2.7}
		\Sigma_{cr}(z,z_s) \equiv \dfrac{M_l}{\sigma(M_l,z,z_s)} = \dfrac{c^2}{4\pi G} \dfrac{D_{os}}{D_{ol} D_{ls}} \gtrsim 10^9 \,\mathrm{M_\odot\, kpc^{-2}}.
	\end{equation}
	
	Therefore, in order to calculate the halo mass $M$ for given $M_l,\,z, \text{and} \,z_s$, we demand a solution of the equation 
	\begin{equation} \label{2.8}
		\Sigma(M;M_l,z) = \Sigma_{cr}(z,z_s), 
	\end{equation}
	where $\Sigma$ is the projected halo surface density described extensively in Sect. \ref{sec3}.
	Solving this equation numerically, we obtain $M(M_l,z,z_s)$. We use the central finite difference approximation to estimate the derivative $dM/dM_l$ which appears in Eq. \eqref{2.6}.     
	
	Given that the halo surface mass density is obtained after an integration of the density profile, it is clear that the results will differ significantly from model to model, since each DM scenario predicts a different density profile.  
	
	\subsection{Expectation value of lensing events}
	Once we obtain the milli-lensing optical depth, we evaluate the expectation number of lensing events in the SMILE source sample using
	\begin{equation} \label{2.9}
		N_l = \sum_{i=1}^{N_{sources}} 1 - \exp\left(-\tau(z_{s,i})\right).
	\end{equation}
	For $\tau \ll 1$, we can approximate Eq. \eqref{2.9} by
	\begin{equation} \label{2.10}
		N_l \approx \sum_{i=1}^{N_{sources}} \tau(z_{s,i}).
	\end{equation} 
	The source sample, as well as their corresponding redshifts are described next.
	
	\subsection{SMILE sample} \label{smile}
	
	The source sample considered in this study is the one of SMILE\footnote{\url{https://smilescience.info/}}: a complete sample built starting from the complete sample used in the Cosmic Lens All-Sky Survey \citep[CLASS;][]{Myers2003,Browne2003}, the most successful search to-date for gravitational lens systems at galactic scales using radio frequencies. The complete sample of 11685 sources presented in CLASS is drawn from two other catalogs: the 5 GHz GB6 catalog \citep{Gregory1996}, and the 1.4 GHz NVSS catalog \citep{Condon1998}. 
	The CLASS catalog contains sources from declination 0$^{\circ}$ to 75$^{\circ}$, with a minimum flux density of 30 mJy at 5 GHz, flat spectral index ($< 0.5$) between 1.4 and 5 GHz, and Galactic latitude ($|b|\geq10^{\circ}$). The complete sample of 11685 sources has been initially followed up in CLASS with low resolution Very Long Array (VLA) observations at 8 GHz. SMILE started from the complete sample in CLASS and selected sources with total flux density at 8 GHz $\geq$ 50 mJy. The 4968 sources that satisfy such a requirement make a complete sample of flat spectrum sources at declination [0$^{\circ}$, +75$^{\circ}$]. 
	
\begin{figure}[h]
		\centering
		\includegraphics[width=9cm]{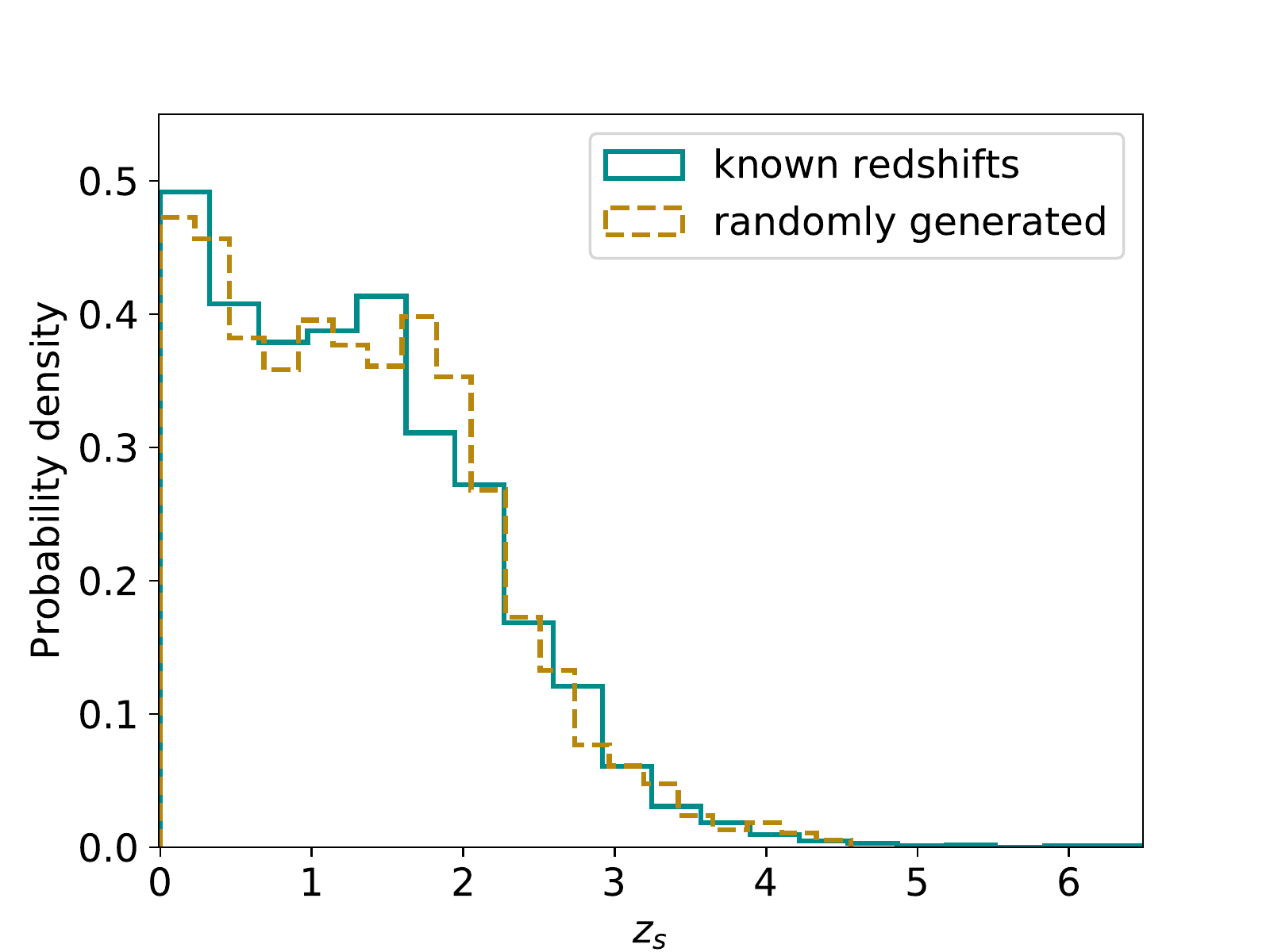}
		\caption{
			Redshift distribution for sources used in this study. Cyan solid line represents the redshift distribution of sources with known redshift, whereas golden dashed line stands for the distribution of the randomly selected redshift measurements from the known redshift sample.  
		}
		\label{fig1}
		%cscript for the derivation of this plot can be found in optical_17_9.ipynb code. 
	\end{figure}

	In order to obtain redshift measurements for sources in the SMILE sample, we used the  Optical Characteristics of Astrometric Radio Sources (OCARS) catalog \citep{Malkin2018}, containing redshift measurements of a large number of radio sources observed in different VLBI astrometry programs. Of the 4968 sources in SMILE, 2781 have an optical counterpart within 3 arcsec with redshift  measurements in OCARS. For the remaining sources we searched for an optical counterpart within 3 arc seconds, with known redshift, in NED\footnote{The NASA/IPAC Extragalactic Database (NED) is funded by the National Aeronautics and Space Administration and operated by the California Institute of Technology.}. In total, we collected redshifts for $\sim2/3$ of sources in SMILE. For the remaining $\sim1/3$, we generated redshift measurements randomly selecting values from the known redshift sample. Their distribution is shown in Fig. \ref{fig1}.
	%The distribution of random redshifts is similar to the one of real redshifts, as can be seen in Fig. \ref{f3}. 
	
	%-------------------------------------------------------------------
	\section{DM haloes \& Mass functions} \label{sec3}
	
	\subsection{Halo size \& structure} 
	The internal structure of dark matter haloes affects the lensing optical depth, since the threshold for strong gravitational lensing is associated with the projected surface mass density which in turn is related to the shape of the density profile. For a detailed review of the various mass densities see \citet{Zavala2019}, while a thorough comparison among various density profiles can be found in \citet{Merritt2006} work.
	
	An important finding of the past decades is that spherically averaged DM density profiles in N-body cosmological simulations have a universal form \citep[][]{Navarro1997}. Such density profiles are described by a simple functional form characterized by only two free parameters. The first one is the concentration parameter, denoted by $c_\Delta$, which quantifies how concentrated the mass is towards the center of the halo. The other one is the characteristic radius, $r_s$, which determines the distance from the center above which the density profile becomes steeper, i.e. quantifies roughly the size of the core. These two parameters are related to each other through
	\begin{equation}
		c_\Delta=\dfrac{r_{\Delta}}{r_s},
		\label{3.4}
	\end{equation}  
	where $r_{\Delta}$ is the virial radius. 
	
	Since the distribution of mass is continuous, the boundary of a halo cannot be defined precisely. So, another major challenge is to come up with a robust method of determining the size of a halo uniquely. Thus far, numerous papers that deal with this problem have been published by several authors \citep{Cole1996,White2001,Cuesta2008,Zavala2019}. In general, the radius of a halo can be defined through the overdensity parameter, $\varDelta(z)$, which in principle depends on the cosmology \citep{Bryan1998,Tinker2008,Naderi2015,Seppi2021}. In particular, it represents the radius where the mean interior density is $\varDelta(z)$ times the critical density of the Universe $\rho_{cr}(z)$, namely
	\begin{equation}
		\dfrac{3}{4\pi r^3_{\Delta}} \int_0^{r_{\Delta}} \rho(\vec r) d^3 \vec{r} = \varDelta \rho_{cr},
		\label{3.5}
	\end{equation}
	where the critical density is given by
	\begin{equation}
		\rho_{cr}(z) = \dfrac{3H^2(z)}{8\pi G} = \rho_{cr,0}\left(\dfrac{H(z)}{H_0}\right)^2,
		\label{3.2}
	\end{equation}
	with $G$ being the Newtonian gravitational constant, while $\rho_{cr,0}$ accounts for the critical density of the Universe at redshift $z=0$, and $H(z)$ is given in Eq. \eqref{3.1}.
	
	The halo mass $M_{\Delta}$, which is the mass contained within a sphere of radius $r_{\Delta}$, is given by
	\begin{equation}
		M_{\Delta} = \varDelta \dfrac{4\pi}{3}r^3_{\Delta} \rho_{cr},
		\label{3.6}
	\end{equation}
	and as a result the halo radius can also be written as
	\begin{equation}
		r_{\Delta}(M_\Delta,z) = \left(\dfrac{3M_\Delta}{4\pi\varDelta\rho_{cr}(z)}\right)^{1/3}.
		\label{3.7}
	\end{equation}
	However, the most commonly used way to determine the halo's size is to consider that the overdensity parameter $\varDelta$ is fixed and equal to 200, since it turns out to be a rather convenient way to define the boundary of the halo and simplifies the calculations (e.g., \citealp{Cole1996}). Taking this fact into account,
	we fix the overdensity to be $\varDelta=200$, throughout this paper, and therefore the halo mass is $M_{200}$ (hereafter, $M$), the halo radius is $r_{200}$, and the concentration is $c_{200}$ (hereafter, c).
	
	Other quantities used extensively below are the enclosed mass, $M_{enc}$, the projected surface mass density, $\Sigma$, and the lens mass, $M_{l}$. The enclosed mass is defined as the mass which is contained within a sphere of radius $r$, namely
	\begin{equation}
		M_{enc}(r) = 4\pi \int_{0}^{r}r'^2 \rho(r') \, dr',
		\label{3.8} 
	\end{equation}
	where we assume that we deal with spherically symmetric objects. The projected surface density is derived simply by the integration of the mass density along the line of sight \citep{Wright2000,Mo2010,Dhar2010,Retana2012, Lapi2012}
	\begin{equation}
		\Sigma(s) = \int_{-\infty}^\infty \rho(\vec r) \,dz, 
		\label{3.9}
	\end{equation}
	where $s=\sqrt{r^2 + z^2}$ is the projected radius (orthogonal to the line-of-sight) relative to the center. 
	
	Having defined the projected surface density we can infer the gravitational lens mass $M_{l}$, by carrying out an integration of the $\Sigma(s)$ over the disk which has radius $s$ \citep[see for example Eq. 41 in][]{Retana2012}
	\begin{equation}
		M_{l}(s) = 2\pi \int_0^s x\Sigma(x)\,dx.
		\label{3.11}
	\end{equation}
	This quantity is the mass contained within an infinite cylinder of radius $s$ in which the mass distribution is characterized by the mass density profile $\rho(\vec r)$. Eq. \eqref{2.8} cannot be solved independently, but must simultaneously satisfy Eq. \eqref{3.11}, owing to the fact that one of the input parameters in Eq. \eqref{2.8} is the lens mass. Therefore, the relation between lens mass and halo mass that is needed in the computation of the lensing optical depth is derived only after solving this coupled non-linear system.   
	%------------------

	\subsection{Mass function} \label{sec3.1}
	
	The computation of the lensing optical depth depends on the number density of lensing objects which in our case are DM haloes. So, we need to obtain a formula that determines the distribution of virialized dark matter haloes (in the field) per volume element and per halo mass at a given redshift z, in order to estimate the lensing optical depth. This problem has been addressed by several authors in the past, either analytically \citep{Press1974,Bond1991,Pavlidou2005} or numerically \citep{Jenkins2001,Tinker2008}. All these works are based on the spherical collapse scenario \citep[e.g.,][]{Gunn1972,Naderi2015}. Improvements using ellipsoidal collapse  do exist \citep[e.g.,][]{Sheth2001}, but we will not use them in this paper.
	
	\subsubsection{CDM mass function} \label{sec3.3.1}
	
	The differential halo mass function of CDM haloes (the number of haloes with mass between the range $M$ and $M+dM$ per proper volume at a given redshift) reads (\citealp{Press1974}, see also Appendix \ref{App.A})
	\begin{equation}
		n(M,z) \equiv\dfrac{dN}{dM} = \sqrt{\dfrac{2}{\pi}} \dfrac{\rho_m(z)}{M}\dfrac{\delta_c(z)}{\sigma_M^2} \left|\dfrac{d\sigma_M}{dM}\right| \exp\left[-\dfrac{\delta^2_c(z)}{2\sigma^2_M}\right],
		\label{3.12}
	\end{equation}
	where $M$ refers to the halo mass, $z$ is the redshift, $\rho_m(z)$ is the mean matter density of the Universe at redshift z, and $\delta_c(z)$ denotes the overdensity of a structure collapsing at redshift z linearly extrapolated to the present. Moreover, $\sigma_M$ is the rms of the density field smoothed on scale M (see Appendix \ref{App.A}).
	
	\subsubsection{WDM mass function}
	Although the halo mass function for WDM haloes shows a similar behavior to the CDM one on galaxy clusters scales, it exhibits a cutoff below the dwarf galaxy scale, owing to the free streaming of WDM particles in the early Universe. The most commonly applied method to derive the WDM halo mass function is the development of N-body simulations that evolve the primordial density field perturbations in time, leading to collapsed DM haloes \citep[e.g.,][]{Schneider2012,Bose2016,Lovell2020a,Lovell2020b}. In this study, we choose to use the numerical fit offered by \citet{Lovell2020b} to take into account the cutoff in the mass function of WDM haloes with respect to the CDM one. The halo mass function in the case of WDM is given by
	
	\begin{equation}
		\label{3.13}
		n_{WDM}(M,z) = n_{CDM}(M,z) \left(1 + \left(\dfrac{\alpha M_{hm}}{M}\right)^\beta\right)^{\gamma},  
	\end{equation}
	where $M_{hm}$ is a characteristic mass scale (the half-mode mass), while $\alpha,\, \beta, \,\gamma$ are parameters of the fit that have been found to be $2.3,\,0.8,\,-1$, respectively. The half-mode mass is associated with the free-streaming length which in turn is related to the rest mass of the WDM particle.
	
	Here, we are interested in exploring the case where the WDM is made of collision-less particles (thermal relics) of mass $m_{WDM} = 3.3$ keV. In this scenario, the theoretical value for the half-mode mass is $M_{hm} \simeq 2 \times 10^8 \, \mathrm{M_\odot}$ \citep[e.g.,][]{Bose2016}. This value for the half-mode mass coincides with the one for the well-motivated sterile-neutrino model in which particles are assumed to have rest mass equal to $7$ keV and lepton asymmetry number $L_6 = 8.66$ \citep[see][]{Bose2016}. Sterile neutrinos are part of the neutrino Minimal Standard Model \citep[$\nu$MSM;][]{Boyarksy2009} which is a simple extension to the Standard Model of particle physics. It has been introduced to explain the unidentified $3.53$ keV X-ray line observed recently in galaxies \citep[see for example,][]{Boyarsky2014} by considering this line to be the decay signal of those $7$ keV sterile neutrinos. 
	
\begin{figure}
		\centering
		\includegraphics[width=9cm]{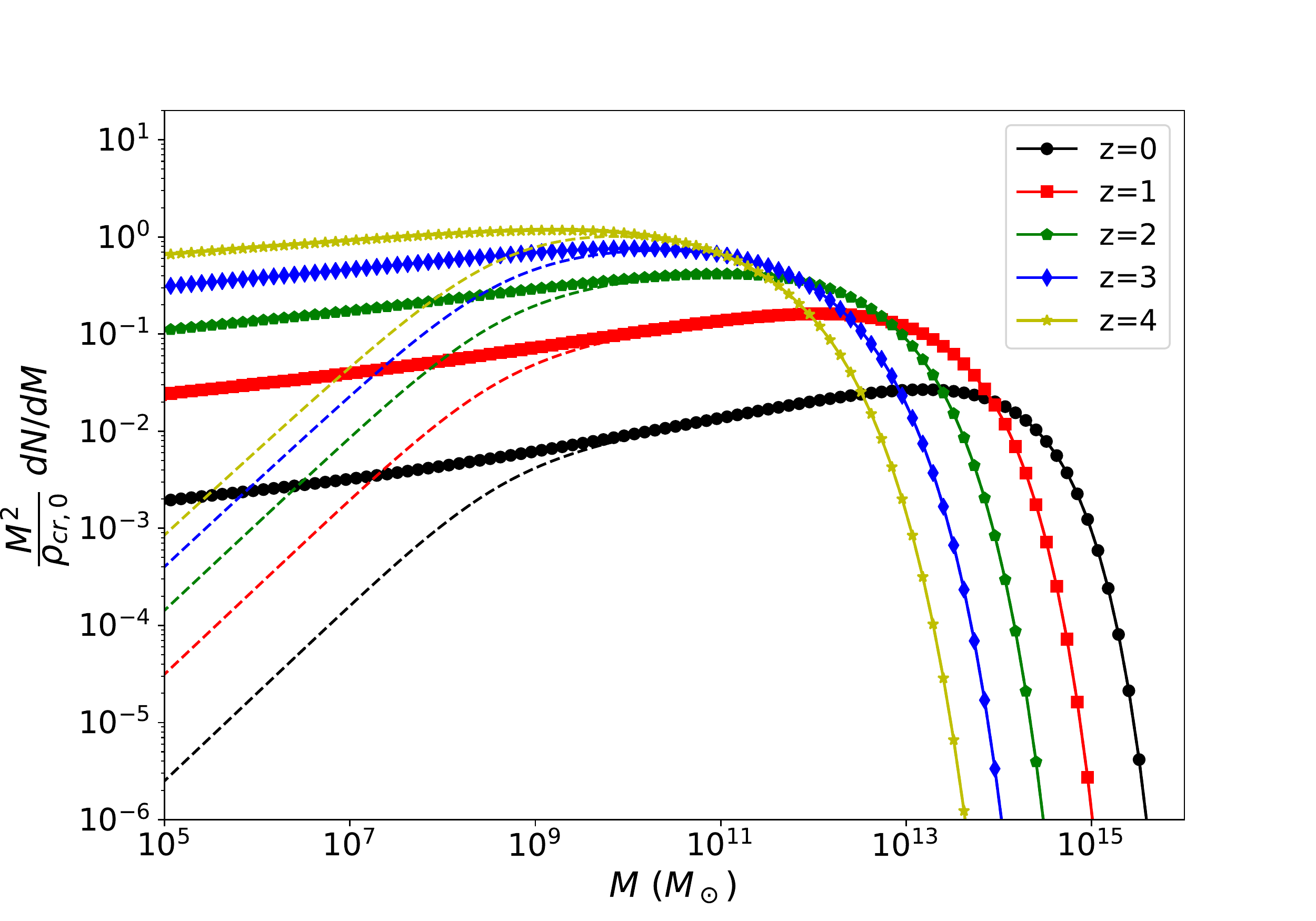}
		\caption{Comparison of the CDM differential halo mass function with the one of the WDM model for various redshifts. Solid lines refer to CDM, while dashed lines correspond to WDM. The vertical axis corresponds to the halo mass while the y-axis shows the proper density of haloes of mass $M$ normalized to the present value of the critical density of the Universe.}
		\label{f2}
	\end{figure}

	Given that the cutoff in the halo mass function of these two WDM models is determined by the same half-mode mass and that the internal structure of haloes is identical, the inferences of this work concerning the $3.3$ keV thermal relic WDM particle will also be valid for the $7$ keV sterile neutrinos model.
	In Fig. \ref{f2}, we display the differential halo mass function for the CDM model and for the WDM one investigated here as a function of the halo mass for various redshifts. Solid lines correspond to the proper density of haloes of mass $M$ at different epochs (redshifts) divided by the present critical density of the Universe, whereas dashed lines represent the same quantities, but for WDM. A major difference between the CDM mass function and the mass function of WDM is that the latter one exhibits a cutoff at the dwarf-galaxy scales ($\sim 10^9~\mathrm{M_{\odot}}$). 	This distinctive feature of the halo mass function in the WDM scenario has a considerable implication to the ability of DM haloes to act efficiently as gravitational milli-lenses on background sources, because the mass function is directly involved in the calculation of the milli-lensing optical depth (Eq. \ref{2.4}).

	\subsection{CDM halo density profile}
	
	In this study, we employ the Navarro-Frenk-White (NFW) profile for the description of the mass distribution within CDM haloes \citep{Navarro1995,Navarro1996,Navarro1997} 
	\begin{equation}
		\rho(r,r_s,c,z)=\rho_{cr}(z)\dfrac{\phi_c}{(r/r_s)\left(1 + r/r_s\right)^2},
		\label{3.18}
	\end{equation}
	where $r_s$ is the characteristic radius, while $\phi_c$ is calculated by
	\begin{equation}
		\phi_c = \dfrac{200}{3} \dfrac{c^3}{\ln(1+c) - c/(1+c)},
		\label{3.19}
	\end{equation}
	with $c$ being the concentration parameter, which is not mass independent, but correlates strongly with the halo mass, as well as with the redshift, following a simple scaling law \citep[see for example,][]{Bullock2001,Neto2007,Prada2012,Dutton2014,Klypin2016,Shan2017,Ragagnin2019,Ragagnin2021}. The NFW profile predicts a cuspy halo's center, since the mass density goes as $\sim r^{-1}$ near the center of the halo.

	Using Eq. \eqref{3.18} along with Eq. \eqref{3.8}, we obtain the enclosed mass
	\begin{equation}
		M_{enc}\left(M,z,r_{200}(M,z)\right) = \dfrac{M}{\ln(1+c) - c/(1+c)} F(r,c,r_{200}),
		\label{3.20}
	\end{equation}
	where
	\begin{equation}
		F(r,c,r_{200}) = \ln(1 + cr/r_{200}) - \dfrac{cr/r_{200}}{1 + cr/r_{200}}. \label{3.21}
	\end{equation}
	From Eqs. \eqref{3.9} and \eqref{3.18}, we obtain the surface density
	\begin{equation}
		\Sigma\left(s,M,z,r_{200}(M,z)\right)=2\dfrac{\phi_c \rho_{cr}(z)r_{200}c^{-1}}{c^2(s/r_{200})^2 - 1 }
		{\cal S}(s, c, r_{200}(M,z)), \label{3.22}
	\end{equation}
	where we have defined for convenience
	\begin{equation}
		{\cal S}\left(s, c, r_{200}(M,z)\right)=	\left\{
		\begin{array}{ll}
			1 - \dfrac{\cos^{-1}(r_{200}/cs )}{\sqrt{c^2(s/r_{200})^2 - 1}} & \mbox{if } s > r_s \\
			1 - \dfrac{\cosh^{-1}(r_{200}/cs)}{\sqrt{1 - c^2(s/r_{200})^2 }} & \mbox{if } s < r_s
		\end{array}
		\right.. \label{3.23}  
	\end{equation}
	From Eq. \eqref{3.11}, the lens mass then is
	\begin{equation}
		M_{lens}\left(s,M,z,r_{200}(M,z)\right)=\dfrac{M}{\ln(1+c) - c/(1+c)}
		{\cal H}(s, c, r_{200}),
		\label{3.24} 
	\end{equation}
	where
	\begin{align}
		{\cal H}(s,c&,r_{200}) = \ln(cs/2r_{200}) \notag \\
		+&\left\{
		\begin{array}{ll}
			\dfrac{2}{\sqrt{(cs/r_{200})^2 - 1}}\arctan{\sqrt{\dfrac{cs/r_{200} - 1}{cs/r_{200}+1}}} & \mbox{if } cs > r_{200} \\
			1 & \mbox{if } cs = r_{200} \\
			\dfrac{2}{\sqrt{1 - (cs/r_{200})^2}}\arctanh{\sqrt{\dfrac{1 - cs/r_{200}}{cs/r_{200}+1}}} & \mbox{if } cs < r_{200}
		\end{array}
		\right. .
		\label{3.25} 
	\end{align}

	\subsection{WDM halo density profile}
	
	WDM is made of particles that had non-negligible thermal velocities at early times. This major difference is expected to have an impact on the concentration of mass near the center, but not in the shape of the distribution of mass within a halo. Indeed, the density profile in WDM models can be well described by a NFW profile \citep[see, e.g.,][]{Lovell2014,Bose2016}. However, DM haloes consisting of WDM are typically formed at smaller redshifts with respect to the formation of CDM haloes. This difference affects the concentration of the halo, which generally reflects the density of the Universe at the epoch of halo formation \citep[for a detailed discussion see,][]{Schneider2012,Bose2016,Zavala2019}. 
	Therefore, we assume that the mass distribution in WDM haloes is consistent with the NFW profile, but more fuzzy, i.e., less concentrated around the center.  In order to calculate the concentration parameter for WDM haloes as a function of the mass and the redshift we use the findings of \citet{Bose2016}. They offer the following simple functional form for the concentration parameter
	\begin{equation}
		\dfrac{c_{WDM}}{c_{CDM}} = \left(1 + \gamma_1 \dfrac{M_{hm}}{M}\right)^{-\gamma_2}  (1+z)^{\beta(z)},
		\label{3.26}
	\end{equation}
	where $\gamma_1 = 60$, $\gamma_2 = 0.17$, and $\beta(z) = 0.026z - 0.04$. $M_{hm}$ is the half-mode mass and in this work is set to be $M_{hm} =2\times 10^8 \, M_\odot$ which corresponds either to the model of thermal relics WDM particles of rest mass $m_{WDM}=3.3$ keV or to the $7$ keV sterile neutrinos model (an extension to the Standard model). Due to their smaller concentrations, WDM haloes will be less likely to exceed the lensing surface-density threshold, resulting in a lower milli-lensing optical depth.

	\subsection{SIDM halo density profile} \label{sec3.6}
	
	SIDM was originally introduced by \citet{Spergel2000} to explain observations of central densities in galaxies within the Local Group. Since then, numerous authors have argued that the self-interaction of particles leads to a core-like profile rather than a cusp-like profile. Such a feature could alleviate the cusp/core problem arising for CDM, and hence is considered a well motivated DM alternative. 
	
	There are two kinds of SIDM theories. In the first case the scattering rate per particle, $\Gamma(r)$, is velocity independent which implies that the ratio of the effective cross section, $\sigma$, to the dark matter particle's mass, m, is constant \citep[e.g.,][]{Rocha2013,Elberts2015}. In the second scenario, the scattering rate is velocity dependent and falls rapidly as the velocity increases \citep[e.g.,][]{Zavala2013}. For a recent discussion on SIDM models, as well as on the observational constraints on the self-scattering cross section, see \citet{Tulin2018}. 
	
	In this work, we will assume that the scattering rate per particle is velocity-independent and has the following form
	\begin{equation}
		\Gamma(r) \propto \rho(r) (\sigma/m) v_{rms}(r),
		\label{3.27}
	\end{equation}
	where $\rho(r)$ is the DM mass density at radius r, while $v_{rms}$ is the rms speed of dark matter particles. We consider a typical value for the ratio $\sigma/m 
	\sim 1~\mathrm{cm^2/g}$, since SIDM models with smaller values, on the order of $0.1~\mathrm{cm^2/g}$, are very similar to the CDM models even on scales smaller than dwarf galaxies and cannot produce detectable deviations from CDM predictions \citep{Zavala2013}. On the other hand, higher values of the cross section per mass, $\sim10~\mathrm{cm^2/g}$, have already been ruled out by cluster observations \citep[see, e.g.,][]{Dawson2012}. 
	
	For the structure of SIDM haloes, we again assume spherical symmetry, but now we use a core-like profile. In fact, 
	the mass density is well approximated by the Burkert profile \citep[see,][]{Burkert1995} which has also two free parameters and is given by the following formula
	\begin{equation}
		\rho_B(r,r_b,\rho_b) 
		= \dfrac{\rho_b}{(1+r/r_b)\left(1+(r/r_b)^2\right)},
		\label{burkert}
	\end{equation}
	where $r_b$ is the scale (core) radius, while $\rho_b$ is the central density. As in the NFW profile, the free parameters of the Burkert profile scale with the halo mass. In the special case where $\sigma/m \sim 1~\mathrm{cm^2/g}$, \citet{Rocha2013} have provided a couple of simple scaling laws that connect both the $r_b$ and $\rho_b$ with the halo mass, using data from N-body simulations. These relations (Eq. 17 and Eq. 20 in \citealp{Rocha2013}) are given below
	\begin{equation}
		\dfrac{r_b}{1 \, \mathrm{kpc}} = 2.21 \left(\dfrac{M_{vir}}{10^{10} \, \mathrm{M_\odot}}\right)^{0.43}, \label{3.29}
	\end{equation}
	\begin{equation}
		\dfrac{\rho_b}{\mathrm{M_\odot/pc^3}} = 0.029 \left(\dfrac{M_{vir}}{10^{10} \, \mathrm{M_\odot}}\right)^{-0.19}. \label{3.30}
	\end{equation}
	
	Since these relations have been derived using the virial mass, $M_{vir}$, instead of $M_{200}$ which we have employed throughout this work, we have to re-scale the density profile to be consistent with Eq. \eqref{3.5}. Eq. \eqref{burkert} can be recast as
	\begin{equation}
		\rho_B(r,M) 
		= {\cal A}(M) \dfrac{\rho_b}{(1+r/r_b)\left(1+(r/r_b)^2\right)},
		\label{3.31}
	\end{equation}
	where 
	\begin{equation}
		{\cal A}(M)= \dfrac{M r^{-3}_b}{2\pi \rho_b} \left[\ln\left(1 + \dfrac{r_{200}}{r_b}\right) + \dfrac{1}{2}\ln\left(1 + \dfrac{r^2_{200}}{r^2_b}\right) - \tan^{-1}\left(\dfrac{r_{200}}{r_b}\right)\right]^{-1},
		\label{3.32}
	\end{equation}
	and now we can use the halo mass $M\equiv M_{200}$ instead of the virial mass in Eqs. \eqref{3.29}, \eqref{3.30}. The term ${\cal A}(M)$ has been derived by requiring the mean  density inside a sphere of radius $r_{200}$ to be $200
	\rho_{cr}$ (see Eq. \ref{3.5}).
	
	Using Eqs. \eqref{3.8} and \eqref{3.31}, we obtain for the enclosed mass
	\begin{equation}
		M_{enc}(r) = {\cal A}(M) \pi \rho_b r^3_b\left[\ln\left(1 +\dfrac{r^2}{r^2_b}\right) + 2\ln\left(1+\dfrac{r}{r_b}\right) - 2\tan^{-1}\left(\dfrac{r}{r_b}\right)\right].
		\label{3.33}
	\end{equation} 
	The surface density cannot in general be derived analytically and therefore we have to perform the integration numerically. In the special case where $s=0$, i.e., for the column density through the line-of-sight, the integration that returns the surface density $\Sigma_B(0)$ yields the closed-form expression 
	\begin{equation}
		\Sigma_B(s=0,M) = {\cal A}(M) \dfrac{\pi}{2}\rho_b r_b,
		\label{3.34}
	\end{equation}
	where the index $B$ indicates that this surface density arises from the Burkert profile. The surface mass density is maximized when $s=0$ since $\rho(r)$ is a monotonically decreasing function of $r$, and as a result for a SIDM halo of given mass, the maximum value of the surface density is determined by Eq. \eqref{3.34}. This feature is of great importance in strong gravitational lensing where the surface density must exceed a critical threshold in order to significantly bend a light ray. 
	
\begin{figure}
		\centering
		\includegraphics[width=9cm]{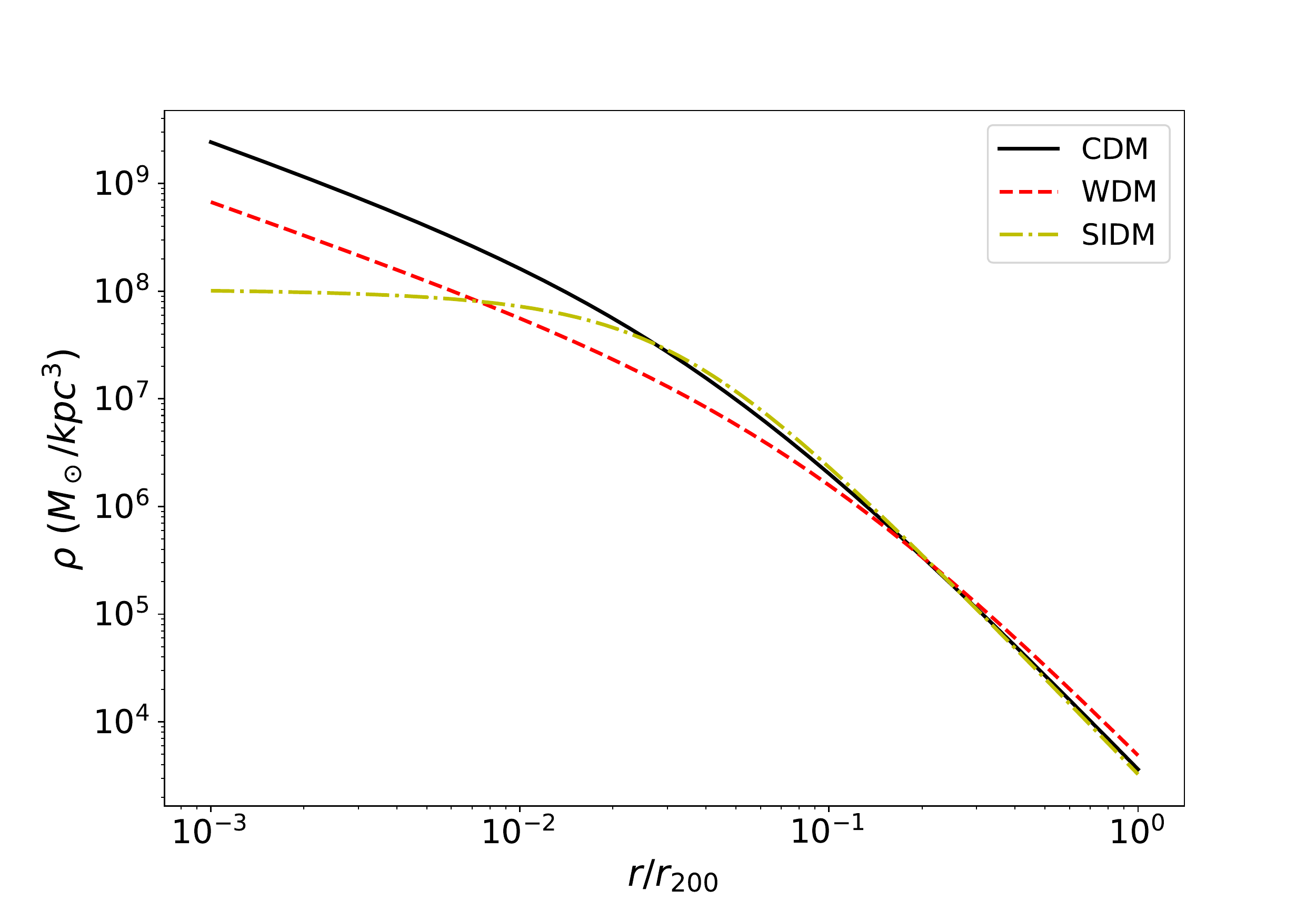}
		\caption{
			Comparison of mass density profiles for a given halo of mass $M=10^8\, \mathrm{M_\odot}$ at redshift $z=0$ for various dark matter scenarios. The concentration for the CDM case is given by Eq. \eqref{concentration}.
			%go to optical_17_9.py code if you wish to print this figure
		}
		\label{f3}
		%cscript for the derivation of this plot can be found in optical_17_9.ipynb code. 
	\end{figure}

	In order to have a qualitative picture of the differences between the three mass density profiles mentioned above, in Fig. \ref{f3}, we apply them to a DM halo of mass $M=10^{8} ~\mathrm{M_\odot}$ at redshift $z=0$.
	For SIDM particles the density profile near the center is flat, while for CDM particles the profile near the center goes as $~r^{-1}$, since we have used the NFW profile. For WDM particles the profile is NFW-like but with larger characteristic radius than in the CDM reflecting the fact that the concentration in the WDM scenario is smaller than the one in CDM.
	
	\subsubsection{SIDM core collapse} \label{sec3.6.1}
	
	Even though most of SIDM models are in favor of a less dense core-like halo center, the strong self-interaction developed between particles in the innermost region of the halo might have important implications in the dynamical evolution of the halo.  
	In one scenario, strong self-interactions between particles induce a negative heat capacity, eventually leading to the formation of a dense central core in the inner part of the halo 
	\citep[see e.g.,][]{Yang2021,Yang2022}. \citet{Yang2021} demonstrated that such a scenario can be successful in explaining the observational excess of small-scale gravitational lenses in galaxy clusters reported in \citet{Meneghetti2020}. 
	They exploited the fact that
	at late stages of the gravothermal evolution of a halo composed of SIDM, the core might undergo gravothermal collapse, resulting in a highly dense halo center, thereby increasing its lensing effect on background sources compared to CDM haloes.
	
	In the most extreme case, the collapsed core can further contract,  eventually leading to the formation of a supermassive black hole (SMBH) at the halo center. This scenario was firstly proposed and studied extensively by \citet{Feng2021} \citep[see also][]{Feng2022} as a possible mechanism to explain the existence and origin of SMBHs at high redshifts ($z
	\sim 6-7$). SIDM offers a natural mechanism for triggering
	dynamical instability, a necessary condition to form a black hole. This scenario can be tested and well-constrained through milli-lensing, since the central SMBH can effectively act as a strong gravitational lens and produce multiple images of a compact background source.
	
	Given that studies dealing with the core collapse scenario do not provide an exact formula for the final mass distribution of DM inside the collapsed halo, we shall restrict ourselves in investigating here only the latter, most extreme, scenario of core collapse where the formation of a SMBH takes place from the gravothermal collapse of the core. The exploration of this model yields an upper limit on the expectation value of lensing events in the SMILE source sample in the case of the SIDM scenario.

	\section{Results} \label{sec4}
	
	\subsection{CDM} \label{sec4.1} 
	
	\subsubsection{CDM: Model A} \label{sec4.1.1}
	We start by investigating the CDM scenario using a concentration-mass relation derived from N-body simulations. We employ the relation given in \citet{Ragagnin2019} to determine the dependence of the concentration parameter $c$ on the redshift $z$, as well as on the halo mass $M$:   
	\begin{equation}
		c(M,z) = 6.02 \left(\dfrac{M}{10^{13}\,\mathrm{M_\odot}}\right)^{-0.12}\left(\dfrac{1.47}{1+z}\right)^{0.16}.
		\label{concentration}
	\end{equation}
	In Fig. \ref{f5} we plot with a blue solid line the milli-lensing optical depth obtained for this c(M,z). The value of the milli-lensing optical depth is well below $\sim 10^{-4}$ implying that even a sample of ten thousands distant ($z\sim5$) compact sources is highly unlikely to produce at least one lensing event. Indeed, performing the summation in Eq. \eqref{2.10} over all sources involved in the SMILE sample, we end up with the value $\left<N_{exp}\right> \simeq 1.5 \times 10^{-3} $, which makes detection of a milli-lens improbable.   
	
	\subsubsection{CDM: Model B} \label{sec4.1.2}
	
	As a limiting case of the possible effect of the concentration-mass relation on our results, we also test a power-law extrapolation to lower masses of the empirical (fitted from observations rather than simulations) c-M relation shown in Fig. 13 of \citet{Prada2012}
	\begin{equation}
		\log c(M,z) = 4.23 - 0.25\log(M/M_\odot) - 0.16\log\left(\dfrac{1+z}{1.47}\right).
		\label{c3}
	\end{equation} 
	Regarding the dependence on redshift, we consider that it is identical to Eq. \eqref{concentration}, but we stress that most studies suggest a weak dependence of the concentration on redshift, so even if we slightly modify the last term in Eq. \eqref{c3} associated with the redshift dependence, the overall results do not change noticeably. In practice, the concentration parameter is set by the halo mass. We note that Eq. \eqref{c3} predicts higher values of the concentration parameter with respect to the ones inferred from N-body simulations. Although this $c-M$ relation has been derived from galaxy cluster observations and might overestimate the $c$ parameter of haloes on sub-galactic scales, recently \cite{Sengul2022} investigated the strong lens system JVAS B1938+666, concluding that sub-galactic DM haloes can be highly concentrated ($c \approx 60$), in line with Eq. \eqref{c3}.
	
Using this relation in Eq. \eqref{2.4}, we obtain the green dash-dotted line in Fig. \ref{f5}, showing the milli-lensing optical depth as a function of the source redshift. Subsequently, using Eq. \eqref{2.10} to compute the expectation value of lensing events in the source sample of SMILE, we obtain $\left<N_{exp}\right> \simeq 1.2$. This value deviates remarkably from the one corresponding to model A (see Sect. \ref{sec4.1.1}), demonstrating that the concentration-mass relation plays a crucial role in the process of strong gravitational milli-lensing and can thus be strongly constrained with milli-lensing observations.  
This value also places an upper limit on the expectation number of detected milli-lenses in the SMILE's source sample, in the case where the properties of DM particles are in line with the framework of the CDM model.
	
\begin{figure}
		\centering
		\includegraphics[width=9cm]{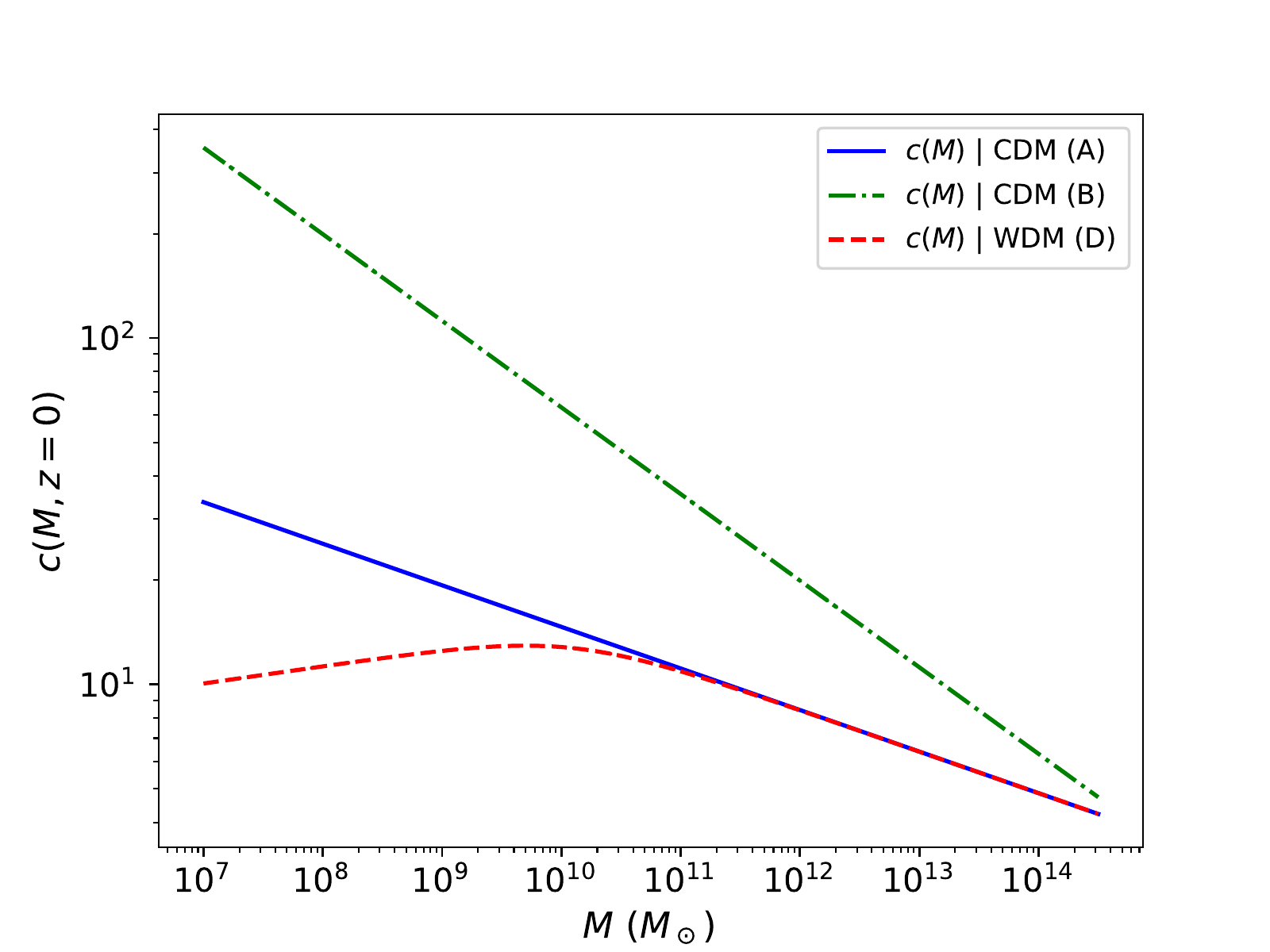}
		\caption{
			Concentration-mass relation at $z=0$ for different dark matter scenarios.
			%Blue solid line represents the (A) CDM scenario in which the concentration parameter, c, is given by Eq. \eqref{concentration}, while green dash-dotted line corresponds to the (B) CDM model in which the c parameter is given by Eq. \eqref{c3}. Red dashed line stands for the (D) WDM model where the $c(M)$ is determined by Eq. \eqref{3.26}.
		}
		\label{f4}
		%a script for the derivation of this plot can be found in the concentration.py file. 
	\end{figure}

	A comparison between the two concentration-mass relations related to the CDM scenario for redshift $z=0$ can be found in Fig. \ref{f4}. The concentration-mass relation given by Eq. \eqref{concentration} (Model A) is displayed with a blue solid line, while the green dash-dotted line stands for the $c(M)$ considered in model B (i.e., Eq. \ref{c3}).
	
	\subsection{SIDM}
	In the SIDM model, there are two possibilities which lead to quite different internal structure of haloes.
	
	The standard scenario is the one where the inner part of haloes is characterized by a core-like profile yielding the projected surface mass density of Eq. \eqref{3.34}. Since the surface mass density is maximized at the center of the halo (as long as the density profile is a decreasing function of $r$), if the central region does not exceed the critical threshold for strong lensing, then the halo will not act as a strong lens. Using Eq. \eqref{3.34} and Eq. \eqref{2.7}, we conclude that the halo mass of a SIDM halo must be $\gtrsim 10^{14}\, \mathrm{M_\odot}$, for the surface mass density at the center to exceed the strong lensing threshold. However, this mass scale corresponds to galaxy clusters and therefore no trustful inferences can be done without including the effect of strong lensing due to the presence of baryons.
	The main finding is that SIDM-only sub-galactic haloes cannot produce milli-lensing images since they are not dense enough to satisfy the strong lensing criterion.
	
	\subsubsection{SIDM core collapse: Model C} \label{sec4.2.1}
	The second scenario related to SIDM haloes is based on the gravothermal core collapse process that might take place in the inner parts of SIDM haloes (see Sect. \ref{sec3.6.1}).
	
	Assuming that the halo in the beginning was described by a NFW profile with the concentration-mass relation to given by Eq. \eqref{concentration}, we can calculate the mass enclosed inside a projected disc with radius equal to the scale radius $r_s$. Then, we consider the most extreme case where the entire core collapses into a very small but extremely dense core that eventually results in the formation of a compact object. This collapsed core is the part of the halo that can produce strong gravitational lensing of light emitted by background sources. 
	
	In Fig. \ref{f5}, we show with the black dotted line the milli-lensing optical depth in the case of SIDM core collapse. Having obtained the optical depth, we carry out the sum shown in Eq. \eqref{2.10} over the redshifts of the SMILE project sources and find the value $\left<N_{exp}\right>\simeq 13$.

	\subsection{WDM: Model D}\label{sec4.3}
	
	In order to derive the milli-lensing optical depth for the scenario of WDM, where particles are supposed to have rest mass $m_{WDM} = 3.3\,\mathrm{keV}$ (thermal relic) or be sterile neutrinos with rest mass equal to $7 \,\mathrm{keV}$, we take into account that the halo mass function is different from that in CDM, and so we adopt the fit offered by \citet{Lovell2020b} (Eq. \ref{3.13}). Regarding the density profile we again employ the NFW one, but with the concentration parameter to be given by Eq. \eqref{3.26}. This concentration-mass relation is however a fit that relates the concentration of WDM haloes with the one corresponding to CDM haloes, so we simply consider that the concentration-mass relation of CDM haloes is the one shown in Eq. \eqref{concentration} and in such a way we obtain a formula for the concentration of WDM haloes as a function of the halo mass and redshift. In Fig. \ref{f4}, we display the concentration-mass relation which corresponds to WDM haloes at $z=0$ with a red dashed line.
	
\begin{figure}
		\centering
		\includegraphics[width=9cm]{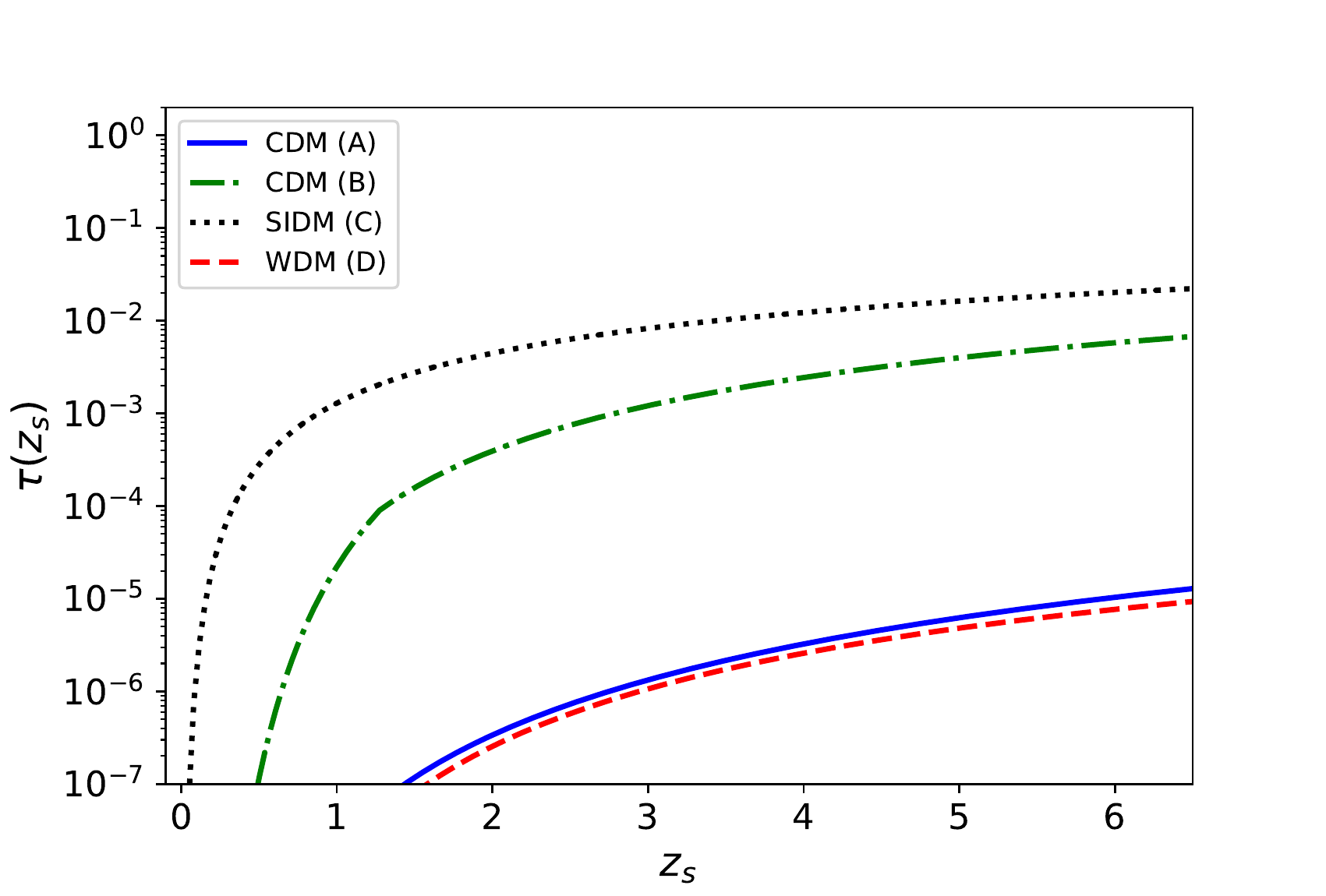}
		\caption{
			Lensing optical depth as a function of the source redshift for different dark matter scenarios.
			%Blue solid line represents the (A) CDM scenario in which the concentration parameter, c, is given by N-body simulations \citep{Ragagnin2019}, while green dash-dotted line corresponds to the (B) CDM model in which the c parameter is given by Eq. \eqref{c3}. Black dotted line stands for the SIDM model which we denote with the letter (C), while the red dashed line refers to the (D) WDM model.
		}
		\label{f5}
		%cscript for the derivation of this plot can be found in optical_depths.ipynb code. 
	\end{figure}

	Performing the integration of Eq. \eqref{2.4}, we find the milli-lensing optical depth in the case of WDM, shown in Fig. \ref{f5} with the red dashed line. Combining this result with Eq. \eqref{2.10}, we compute the expectation number of detected WDM milli-lenses, obtaining $\left< N_{exp}\right> \simeq 1.1 \times 10^{-3}$. It is therefore extremely unlikely to detect any milli-lenses with SMILE if DM is in the form of WDM.

	\section{Discussion \& Conclusions} \label{sec5}
	
	In this work we have explored the ability of sub-galactic DM haloes to act as milli-lenses on background sources resulting in multiple images of the same source with angular separation on the order of milli-arcseconds, considering different DM models. We have developed a semi-analytical method to estimate the expectation value of detected milli-lenses in several DM scenarios, computing the lensing optical depth. 
	We have modeled the number density and internal structure of haloes using either (semi)analytical calculations or fits to N-body simulation results, depending on the DM model. We have restricted ourselves in applying the point-mass lens approximation to infer the lens mass, imposing the effective surface threshold criterion for strong lensing to connect the lens mass to the halo mass. Finally, we used the milli-lensing optical depth in each scenario to calculate the expectation number of detected milli-lenses in the source sample of the SMILE project.
	
	We found that the probability of strong milli-lensing by DM haloes strongly depends on the model, being regulated by the properties of DM particles which dictate the inner structure of haloes, as well as their number density. We have shown that even within the CDM model, the lensing optical depth is quite sensitive to variations in the concentration-mass relation, leading to very different expectation values of detected milli-lenses in the SMILE source sample. Milli-lensing observations might therefore enable us to constrain the concentration-mass relation down to sub-galactic mass scales.
	
	In addition, we have demonstrated that DM scenarios which are in favor of core-like density profiles, such as the SIDM one investigated here, are unlikely to produce milli-lenses because they predict haloes with low-density centers. However, our method allows to also probe scenarios like core collapse which enhance considerably the probability of milli-lensing.
	
	Finally, we have shown that haloes consisting of WDM lead to an extremely small milli-lensing optical depth due to their combination of low concentration and mass-function cutoff. Even if a steeper concentration-mass relation (such as Eq. \ref{c3}) is used, the cutoff in the number density below sub-galactic scales still prevent the milli-lensing optical depth from increasing significantly. Therefore, the detection of milli-lenses would provide definitive evidence against the WDM model and more generally models that exhibit a cutoff in their halo mass function affecting the $10^6 - 10^9 ~\mathrm{M_\odot}$ mass scales.
	
\begin{figure}
		\centering
		\includegraphics[width=9cm]{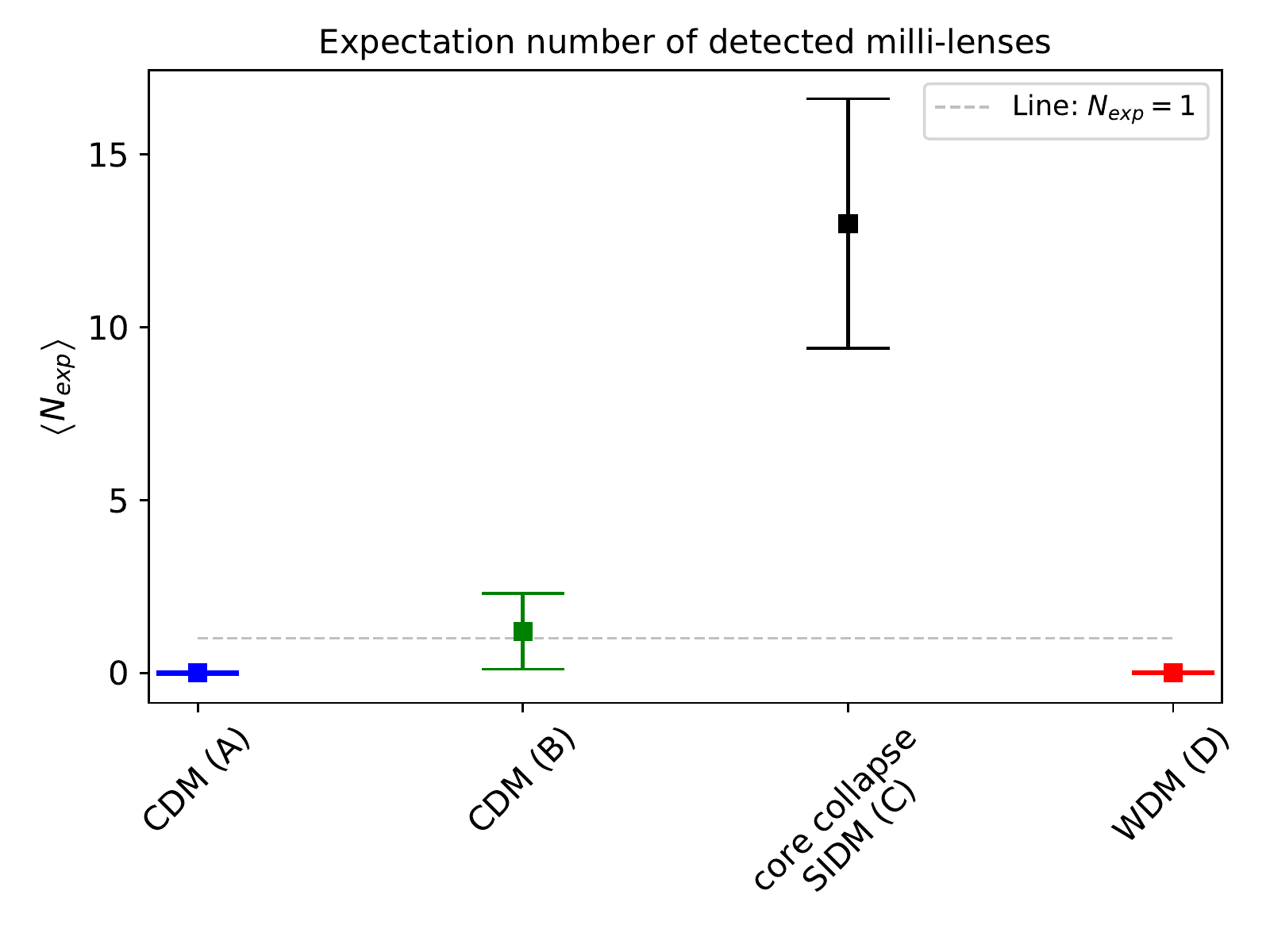}
		\caption{
			Expectation number of detected milli-lenses in the SMILE project.
			%Blue point refers to the (A) CDM scenario, green point accounts for the (B) CDM model, consisting an upper limit on the value of detected milli-lenses in the case of CDM. Black point represents the value of expected milli-lensing events in the case of SIDM (model C), and red point corresponds to the (D) WDM model. 
			The error-bars have been derived assuming Poisson-like error in the calculation of the expectation values.
		}
		\label{fig6}
		%cscript for the derivation of this plot can be found in exp_val.py code. 
	\end{figure}

	In Fig. \ref{fig6}, we summarize our results, plotting the expectation value of detected milli-lenses for all models investigated in this study. The blue point corresponds to the CDM model A (Sect. \ref{sec4.1.1}), while the green point refers to the CDM model B (Sect. \ref{sec4.1.2}). The black point corresponds to the core collapse SIDM scenario C (Sect. \ref{sec4.2.1}) and the red point to the WDM model D (Sect. \ref{sec4.3}). Even among the limited number of DM models studied here, milli-lensing observations of source samples comparable to that of SMILE hold significant discriminating power.

	Even though sub-galactic haloes are expected to be almost empty of baryons, one source of uncertainty in our work might be the fact that we ignore the overall effect of baryons in the internal structure of haloes which in principle can alter the ability of those haloes to act as milli-lenses. Although such systems are DM dominated, the baryons might play a crucial role in the strong lensing and therefore it is left for a future investigation, since the purpose of this paper is mostly to highlight the point that milli-lensing observations can be used to constrain the nature of dark matter and further discriminate between currently viable models.
	
	Another possible source of uncertainty arises from the fact that some DM haloes might host a SMBH at their center, which would contribute significantly in the strong lensing signal from those haloes, thereby modifying our results. However, neither the fraction of haloes that host such objects at their centers nor the accurate relation between SMBH and halo mass are known. Hence we do not take this possibility into account in this study, but we plan to explore this scenario in the future.  
	
	Here, we assumed a certain redshift distribution for the $\sim1/3$ of sources for which no redshift measurements were currently available. This choice partially affects the results, although not our qualitative conclusions. For instance, if we instead assumed that missing redshifts are due to weakest and farthest sources, then the real redshift distribution for these sources would extend toward higher redshifts which correspond to higher values of the optical depth, thereby increasing the expectation number of detected milli-lenses and enhancing the probability of detecting a milli-lens.
	
	In this study, we have concluded that: 1) the source sample included in the SMILE project is sufficiently large to enable inferences about the nature of DM; 2) WDM haloes are highly unlikely to produce even a single strong milli-lensing event in the source sample of SMILE; 3) SIDM haloes can only act as strong milli-lenses in the case where self-interactions trigger the core collapse mechanism, leading to highly dense cores; and 4) the ability of CDM sub-galactic haloes to act as milli-lenses strongly depends on the mass-concentration relation. Finally, we have shown that if CDM is indeed the relevant model for describing the properties of DM particles, then milli-lensing observations will enable us to further constrain the relationship between concentration and halo mass down to sub-galactic mass scales.

	\begin{acknowledgements}
		NL would like to thank Hai-Bo Yu for fruitful discussions and comments related to the core collapse scenario in the context of SIDM. NL and KT acknowledge support by the European Research Council (ERC) under the European Unions Horizon 2020 research and innovation programme under grant agreement No. 771282. VP acknowledges support by the Hellenic Foundation for Research and Innovation (H.F.R.I.) under the “First Call for H.F.R.I. Research Projects to support Faculty members and Researchers and the procurement of high-cost research equipment grant” (Project 1552 CIRCE), and by the Foundation of Research and Technology - Hellas Synergy Grants Program through project MagMASim, jointly implemented by the Institute of Astrophysics and the Institute of Applied and Computational Mathematics. CC acknowledges support by the European Research Council (ERC) under
		the HORIZON ERC Grants 2021 programme under grant agreement No. 101040021. KT acknowledges support from the Foundation of Research and Technology - Hellas Synergy
		Grants Program through project POLAR, jointly implemented by the Institute of
		Astrophysics and the Institute of Computer Science.
	\end{acknowledgements}

	\bibliography{references}

	\begin{appendix}
		\section{Press-Schechter mass function} \label{App.A}
		In this appendix, we offer a short prescription for calculating the halo mass function. For more details on the subject, see \cite{Mo2010}.
		
		Using the Press-Schechter formalism \citep{Press1974}, the differential halo mass function is expressed as
		\begin{equation}
			n(M,z) \equiv \dfrac{dN}{dM} = \sqrt{\dfrac{2}{\pi}} \dfrac{\rho_m(z)}{M}\dfrac{\delta_c(z)}{\sigma_M^2} \left|\dfrac{d\sigma_M}{dM}\right| \exp\left[-\dfrac{\delta^2_c(z)}{2\sigma^2_M}\right], \label{A1}
		\end{equation}
		where $\rho_m(z) = \Omega_m (1+z)^3$, $\delta_c(z)$ is the overdensity of a structure collapsing at redshift z linearly extrapolated to the present epoch, and $\sigma_M$ accounts for the linear rms fluctuation (variance) of the density field on scale M.
		Under the spherical collapse assumption and concordance cosmology, the critical overdensity is given by (e.g., Eq. C30 in \citealp{Pavlidou2005})
		\begin{equation} \label{A2}
			\delta_c(z) = \dfrac{1}{D(z)} \delta_c(0),  
		\end{equation}
		where $D(z)$ is the normalized linear growth factor, i.e., $D(z=0)=1$, calculated by
		\begin{equation} \label{A3}
			D(z) = G\left[(2\omega)^{1/3}/(1+z)\right] \big / G\left[(2\omega)^{1/3}\right],
		\end{equation}
		while $\omega = \Omega_\Lambda / \Omega_m$ and
		\begin{equation} \label{A4}
			G(u) =\dfrac{(2+u^3)^{1/2}}{u^{3/2}} \int_0^u \dfrac{y^{3/2}}{(2+y^3)^{3/2}} dy.
		\end{equation}
		
		The variance $\sigma_M$, normalized to be equal to $\sigma_8$ when $R=8 h^{-1} \, \mathrm{Mpc}$, is given by (see Eq. 10 in \citealp{Pavlidou2005})
		\begin{equation} \label{A5}
			\sigma^2_M = \sigma^2_8 \dfrac{\int_0^\infty dk \,\, P(k) W^2(kR(M)) k^2}{\int_0^\infty dk \,\,P(k) W^2(k8h^{-1}\, \mathrm{Mpc}) k^2},
		\end{equation}
		where the $\sigma_8$ parameter is a direct observable quantity, while $W(kR)$ refers to the window function (filter) and $R$ is the characteristic radius of the filter related to the mass M through
		\begin{equation}
			R(M)=\left(\dfrac{M}{\gamma_f \rho_m}\right)^{1/3}, \label{A6}
		\end{equation} 
		where $\gamma_f$ is a parameter depending on the shape of the filter. Here, we employ a sharp in k-space filter given by
		\begin{equation}
			W(\vec k;R) = 
			\left\{
			\begin{array}{ll}
				1& \mbox{if } |\vec k| \le R^{-1} \\
				0 & \mbox{if } |\vec k| > R^{-1}
			\end{array}
			\right. \label{A7}.
		\end{equation}
		Given this choice, $\gamma_f$ becomes equal to $6\pi^2$ \citep{Mo2010}. 
		
		In Eq. \eqref{A5}, P(k) stands for the linear matter power spectrum. Invoking linear theory, we can write the matter power spectrum to be
		\begin{equation} \label{A8}
			P(k) \propto T^2(k) P_{init}(k),
		\end{equation}
		where $P_{init}$ is the initial power spectrum proportional to $k^n$ with $n = 0.97$, while $T(k)$ corresponds to the transfer function. \citet{Bond1984} offer the following simple numerical formula for the transfer function $T(k)$ which is consistent with the $\Lambda$CDM model (see also \citealp{Jenkins2001})
		\begin{equation} \label{A9}
			T(k) =\dfrac{1}{\left[1 + \left[aq + (bq)^{3/2} + (cq)^2\right]^\nu\right]^{1/\nu}},
		\end{equation}
		where  $q =k/\Gamma$, $\Gamma = \Omega_{m,0} h$, $\nu=1.13$, $a=6.4~ h^{-1} ~ \mathrm{Mpc}$, $b=3 ~ h^{-1} \, \mathrm{Mpc}$, and $c=1.7 ~ h^{-1} \, \mathrm{Mpc}$.

	\end{appendix}
\end{document}